\pdfoutput=1
\documentclass[11pt]{article}

\usepackage[]{acl}

\usepackage{times}
\usepackage{latexsym}

\usepackage[T1]{fontenc}
\usepackage[utf8]{inputenc}

\usepackage{microtype}

\usepackage{inconsolata}

\title{ProductAgent: Benchmarking Conversational Product Search Agent\\with  Asking Clarification Questions}

\usepackage{amsmath}            % 数学公式核心宏包
\usepackage{amssymb}            % 提供丰富的数学符号
\usepackage{amsfonts}           % 提供丰富的数学符号
\usepackage{multirow}           % 表格
\usepackage{booktabs}           % 表格
\usepackage{makecell}           % 表格
\usepackage{tabularx}           % 表格
\usepackage{arydshln}
\usepackage{graphicx}           % 插入图片的宏包
\usepackage{float}              % 设置图片浮动位置的宏包
\usepackage{bm}                 % 字体加粗
\usepackage[tikz]{bclogo}       % Box
\usepackage{caption}

\definecolor{brickred}{HTML}{b92622}
\definecolor{midnightblue}{HTML}{005c7f}
\definecolor{salmon}{HTML}{f1958d}
\definecolor{burntorange}{HTML}{f19249}
\definecolor{junglegreen}{HTML}{4dae9d}
\definecolor{forestgreen}{HTML}{499c5e}
\definecolor{pinegreen}{HTML}{3d8a75}
\definecolor{seagreen}{HTML}{6bc1a2}
\definecolor{limegreen}{HTML}{97c65a}

\newcommand{\Agent}{ProductAgent}
\newcommand{\Benchmark}{\textsc{ProClare}}

\author{
Jingheng Ye$^{1}$,
Yong Jiang$^{2}$\thanks{Corresponding author: Yong Jiang and Hai-Tao Zheng. (E-mail: yongjiang.jy@alibaba-inc.com, zheng.haitao@sz.tsinghua.edu.cn)},
Xiaobin Wang$^{2}$,
Yinghui Li$^{1}$      \\
\textbf{Yangning Li}$^{1,3}$,
\textbf{Hai-Tao Zheng}$^{1,3*}$,
\textbf{Pengjun Xie}$^{2}$,
\textbf{Fei Huang}$^{2}$  \\
$^{1}$Tsinghua Shenzhen International Graduate School, Tsinghua University  \\ 
$^{2}$DAMO Academy, Alibaba Group,
$^{3}$Peng Cheng Laboratory  \\
\texttt{yejh22@mails.tsinghua.edu.cn}  \\
\texttt{\{yongjiang.jy,xuanjie.wxb\}@alibaba-inc.com}
}

\begin{document}
\maketitle

\begin{abstract}

This paper introduces the task of \textit{product demand clarification} within an e-commercial scenario, where the user commences the conversation with ambiguous queries and the task-oriented agent is designed to achieve more accurate and tailored product searching by asking clarification questions.
To address this task, we propose\textbf{~\Agent}, a conversational information seeking agent equipped with abilities of strategic clarification question generation and dynamic product retrieval. Specifically, we develop the agent with strategies for product feature summarization, query generation, and product retrieval.
Furthermore, we propose the benchmark called \textbf{~\Benchmark} to evaluate the agent’s performance both automatically and qualitatively with the aid of a LLM-driven user simulator.
Experiments show that~\Agent~interacts positively with the user and enhances retrieval performance with increasing dialogue turns, where user demands become gradually more explicit and detailed.\footnote{All the source codes will be released after the review anonymity period.}

\end{abstract}
\section{Introduction}\label{sec:introduction}

\begin{table}[tb!]
    \renewcommand{\tabcolsep}{0pt}
    \renewcommand{\arraystretch}{1.0}
    \centering\small
    \resizebox{0.472\textwidth}{!}{
        \begin{tabularx}{\linewidth}{X}
        \toprule
        \textbf{A conversation example}  \\
        \midrule
            \textbf{A}: I am a product sales assistant. Please just type in the targeted product category first, and I would search for you. \\ \\
            \textbf{U}: Sports shoes.  \\ \\
            
            \textbf{A}: I have search several relevant items for you: <Item$_{1}$>, <Item$_{2}$>, $\cdots$. Also, you can answer these questions so that I could recommend more suitable items for you! \\
            
            \textbf{Q1}: What is the suitable scenario for the sports shoes you want to purchase? \\
            \textbf{C1}: ["Outdoor", "Brisk walking", "Basketball", "Dance", "Travel", "Other"] \\
            
            \textbf{Q2}: What style of sports shoes do you like? \\
            \textbf{C2}: ["Versatile", "Internet celebrity", "Foreign style", "Basketball", "Breathable style", "Other"] \\
            
            \textbf{Q3}: Please choose your preferred color for sports shoes. \\
            \textbf{C3}: ["Light gray", "White gray", "Moonlight", "Yellow", "Rose Red", "Other"] \\ \\

            \textbf{U}: \textbf{Answers for Q1}: "Outdoor", "Basketball". \\
            \textbf{Answers for Q2}: "Basketball", "Breathable style". \\
            \textbf{Answers for Q3}: I like green. \\ \\

            \textbf{A}: I have search several relevant items for you: <Item$_{1}$>, <Item$_{2}$>, $\cdots$. Also, you can answer these questions... \\
        \bottomrule
        \end{tabularx}
    }
\caption{A conversation example between~\Agent~(A) and the user (U). Since the clarification questions are formulated as multi-choice questions, the user can conveniently respond by selecting the preferred options or typing in the answers when needed.}
\label{tab:example}
\end{table}

For decades, researchers have been attempting to develop agents with satisfying abilities of understanding human intentions and making decisions~\cite{wooldridge1995intelligent,agha1986actors}. However, these efforts have often been limited to constrained situations~\cite{aliannejadi-etal-2021-building}, where user-system interactions are insufficient~\cite{vats2024survey}. This limitation becomes more prominent and even critical especially in high-stakes domains or task-oriented systems~\cite{yi2024survey}, where users often struggle to express their ambiguous and personal information requirements in the first place~\cite{rahmani-etal-2023-survey}.

% When encountering intractable real-world queries, previous studies tend to enhance the abilities of agents in reasoning and planning~\cite{valmeekam2024planning}, regardless of the ambiguity of input queries. The reasoning and planning power of agents, however, if not guided properly by human involvement, can become uncontrolled and even cause unexpected and even counterproductive outcomes~\cite{ji2023survey}.

The focus of this paper is e-commerce product search, where users usually spend excessive time in locating their desired products since they might not be familiar about which product aspects are suited to their individual use case. We claim that incorporating conversational information seeking could bridge the gaps between traditional product search and offline shopping, where expert salesmen can provide support in identifying the target products that are suitable to users~\cite{papenmeier2022mhm}.
Therefore, we propose the task of \textit{product demand clarification} as illustrated in Table~\ref{tab:example}, which aims to investigate the role of \textit{conversational information seeking} (CIS) in e-commerce product search~\cite{zhang2018towards}. Specifically, in each turn, the agent retrieves relevant product items and asks clarification questions~\cite{lee-etal-2023-asking} to refine user requirements. As the interaction progresses, it is expected that the user demands become clearer, leading to more accurate product searching. This task is challenging since it necessitates the agent to possess abilities of accessing extensive product knowledge~\cite{peng2024ecellm,li2023ecomgpt}, understanding subtle user intents, asking strategic clarification question, and retrieving relevant product items. Additionally, products often include different facets, such as brand, color, and style, resulting in different preference slot combinations or shopping schema~\cite{yang2018query}.

\begin{figure}[tbp!]
\centering
\includegraphics[scale=0.27]{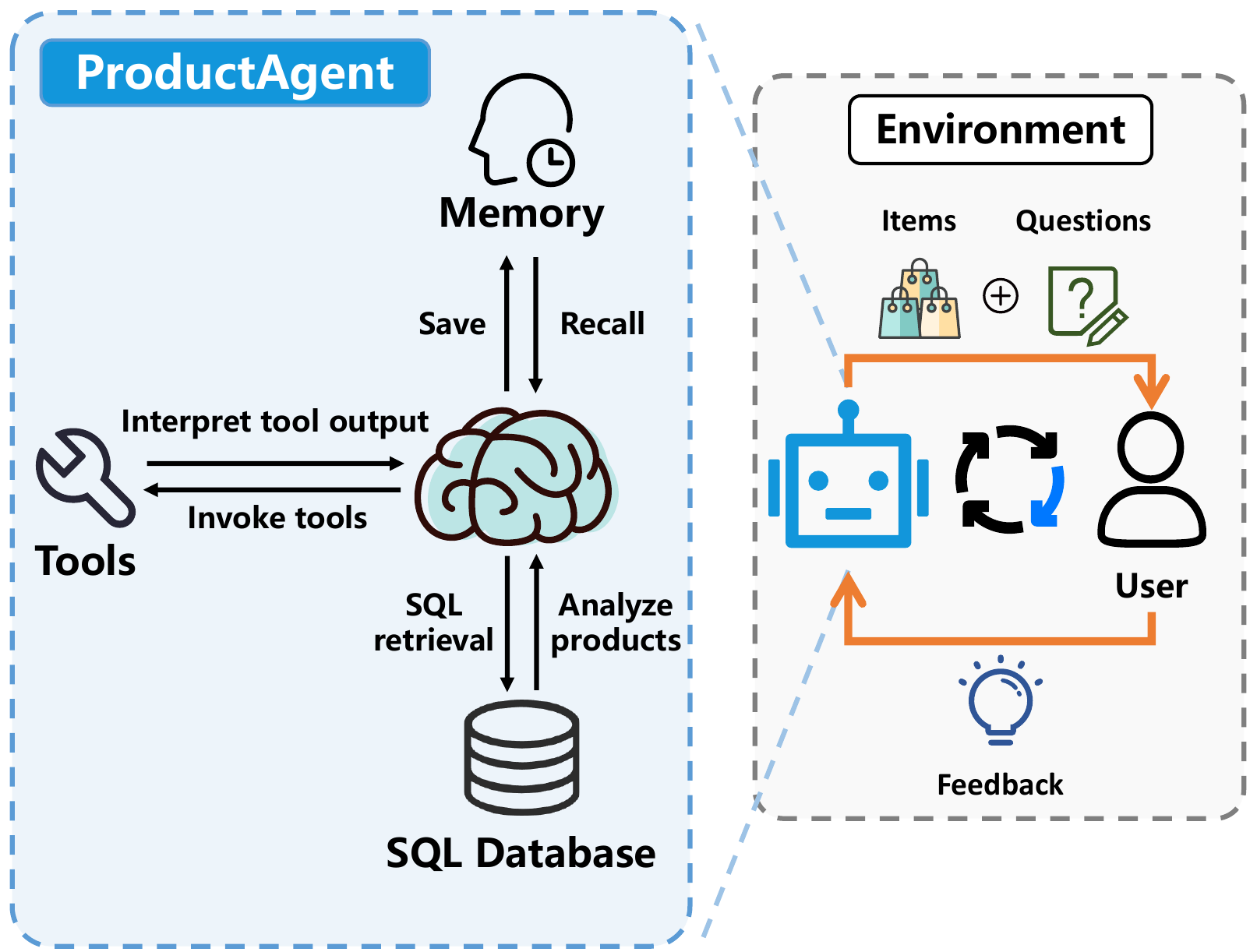}
\caption{Simplified illustration of our~\Agent.}
\label{fig:intro}
\end{figure}

To address this task, We propose\textbf{~\Agent}, a conversational information seeking agent that can serve as a virtual shopping assistant. Technically,~\Agent~performs autonomous loops that integrate three core modules: product databases, a memory module, and a set of tools (Figure~\ref{fig:intro}). Specifically, 1) the databases store product items in both structured and vectorized forms, enabling~\Agent~to retrieve relevant products and summarize them into statistics reflective of the current user demands. 2) The memory module caches a user’s contextual information throughout a conversation session including structured clarification questions and unstructured dialogue records, enabling the agent to ask next questions dynamically. 3) The available tools support various actions such as product retrieval and clarification question generation. By leveraging carefully designed prompts, we enable effective interaction between language and machine learning models (LLMs) and these modules, automating the task.

In order to comprehensively evaluate~\Agent, we introduce the~\textbf{PROduct CLArifying REtrieval} (\textbf{\Benchmark}) Benchmark, which focuses on automatic conversational retrieval evaluation~\cite{zhou2023webarena,semnani2023wikichat}. This evaluation is implemented using the user simulation technique~\cite{zhu-etal-2020-convlab,yoon2024evaluating}, in which a user simulator driven by a LLM~\cite{terragni2023context,sekulic2024reliable} makes it convenient and efficient to generate simulated user responses, thus eliminating the need for human utterances. Specifically, we synthesize a dataset consisting of 2,000 dialogues between~\Agent~and the user simulator, with each dialogue centered around a specific product item of the user's interest. Using this dataset, we build an automatic and reliable retrieval evaluation pipeline for the product demand clarification task. Our experiments in this paper showcase the significant effectiveness of~\Agent, as the retrieval performance improves with the number of interaction turns. Our contributions are as follows:

\begin{itemize}
    \item [(1)] We formulate the task of product demand clarification and propose the solution called~\Agent, which aims to pinpoint user demands by asking clarification questions in a conversational manner.

    \item [(2)] We introduce an straightforward benchmark, called~\Benchmark~, to ensure a reliable and automatic evaluation pipeline for the product demand clarification task.

    \item [(3)] Extensive experiments are conducted to confirm the effectiveness of our approach. Extra extensive analyses are also carried out to reveal promising improvements for the agents.
\end{itemize}

\section{Related Work}\label{sec:related_work}

\subsection{LLM-based Agents}
Building agents centered on LLMs has been widely studied recently~\cite{park2023generative,liu2023training,nakano2021webgpt}, owing to their preeminent abilities of reasoning~\cite{yao2023react,lu2024chameleon}, planning~\cite{song2023llm,huang-etal-2023-frustratingly}, memorizing, reflexion~\cite{shinn2024reflexion,huang2024queryagent} or self-correction~\cite{gou2023critic}, and tool utilization~\cite{schick2024toolformer,qin2023tool}.
Thanks to techniques like Chain-of-Thought (CoT)~\cite{wei2022chain,kojima2022large,wang2022self} and problem decomposition, LLM-based agents are able to solve challenging problems across a wide rage of real-world scenarios~\cite{li2023camel}.
However, many advanced agents focus on resolving real-world problems with minimal human intervention. This often results in bias~\cite{ye-etal-2023-cleme} and the lack of flexibility and inefficiency, particularly in task-oriented scenarios that emphasize interaction~\cite{wang2023voyager}.
The involvement of human feedback~\cite{peng2023check,zhang2024training}, therefore, becomes progressively critical in guiding and supervising the action trajectory of agents, ensuring their alignment with human requirements and objectives~\cite{kenton2021alignment,ngo2022alignment}.

\subsection{Asking Clarification Questions}
Aiming to provide task-specific services, task-oriented agents often confront vague queries from users, who may lack the required domain knowledge to articulate their queries accurately. The ensuing reduced performance is mitigated in recent studies~\cite{zamani2020generating,kuhn2022clam}, which advocate to ask clarification questions to pinpoint user demands in a conversational manner. 
For example, MAS2S~\cite{feng2023towards}introduces the idea of asking questions based on user profiles and task knowledge to clarify user requests.
\citet{lee-etal-2023-asking} presents the CAMBIGNQ dataset consisting of ambiguous questions. This dataset provides a foundation for a pipeline system enabling the asking of clarification questions to become more efficient.
Furthermore, recent studies~\cite{wu2024large,erbacher2023circle,zhang2024clamber} applies clarification questions onto LLMs, achieving significant improvements across various downstream applications.
A concurrent study~\cite{chang2024injecting} develops SalesAgent that concentrates on injecting dialogue strategies into LLMs by fine-tuning, which is orthogonal to our work. Additionally,~\citet{vedula2024question} also propose to generate product-related clarification questions through prompting and fine-tuning LLMs.

However, all these works rely heavily on the internal knowledge of LLMs to generate clarification questions, often compromising the quality of the generated clarification questions. On the contrary, our work leverages specific databases and external tools, thus unburdening LLMs from knowledge acquirement and promoting the generation of quality clarification questions.

% Furthermore,~\citet{wu2024large} applies clarification questions to LLM-based code generation by proposing a communication-centric process, which allows for refined code based on user responses. 
% However, they do \textit{not} take advantage of advanced techniques such as Retrieval-Augmented Generation (RAG) or external tools, which could significantly improve the quality of the generated questions.

\begin{figure*}[tbp!]
\centering
\includegraphics[scale=0.28]{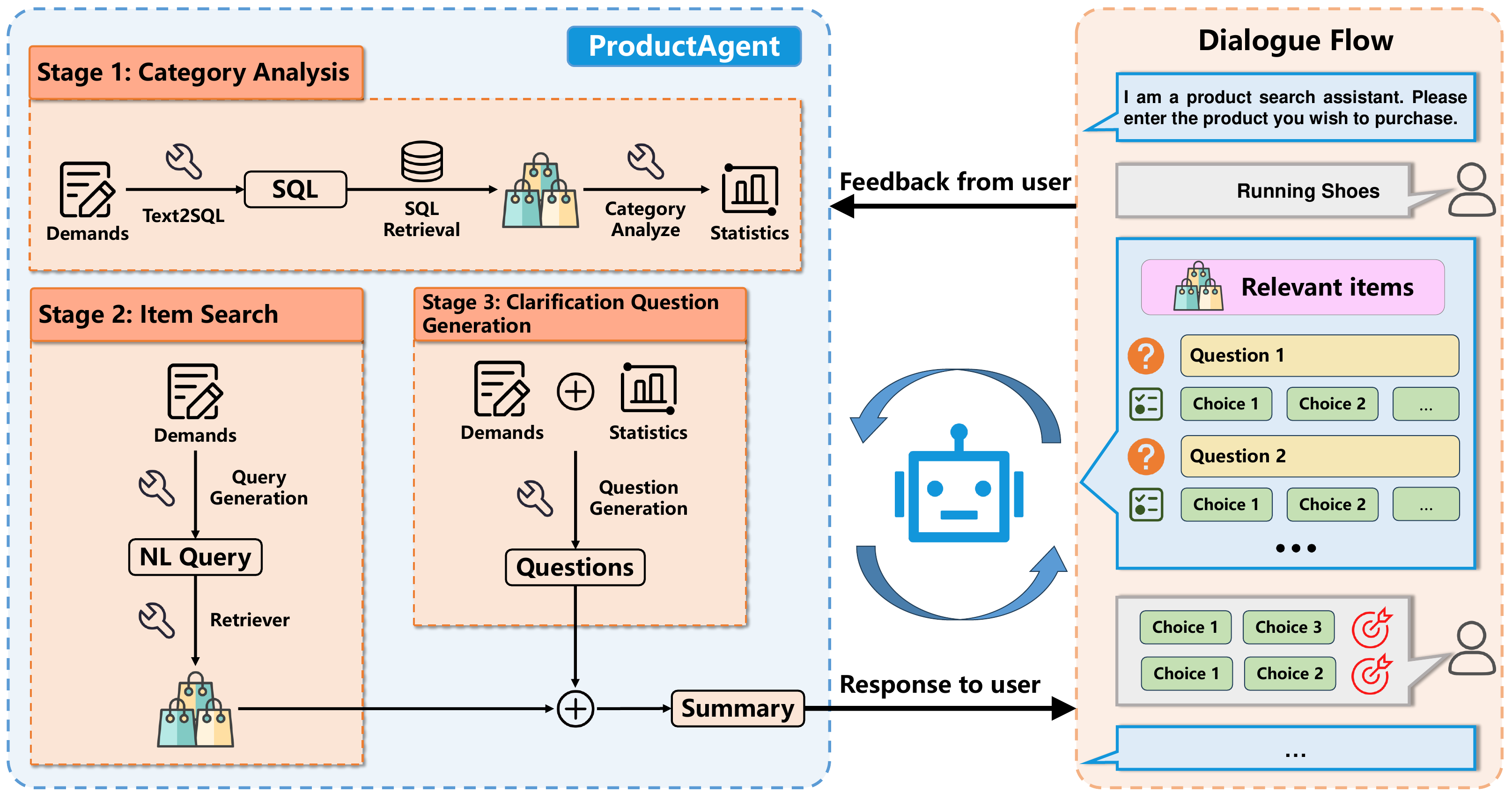}
\caption{Overview of our proposed~\Agent.}
\label{fig:overview}
\end{figure*}

\section{Product Demand Clarification Task}\label{sec:task}

The objective of the product demand clarification task is to enhance the e-commerce search process through the collaboration between humans and the agent. This task is based on a typical online shopping scenario that users (customers) have decided to purchase a product of a certain category but they have not yet determined detailed aspects of the product of their interests.
So they often struggle to clearly articulate their purchasing requirements, especially for product categories accompanied by unfamiliar jargon. To tackle this challenge, the task aims to identify user demands by explicitly posing clarification questions that help clarify any uncertainties towards product searching.

The task takes the form of conversation, where an agent proactively interacts with a user. Specifically, given a product category as $U_1$, the conversation $D$ between them is formulated as follows:

\begin{equation}\begin{aligned}
    D = \{U_1,A_1,P_1, U_2,A_2,P_2, \cdots\}
\end{aligned}\end{equation}
where $U_t$ $(t=1,2,\cdots)$ is the user utterance, $A_t$ is the agent utterance, and $P_t$ is the retrieved product items. The agent utterance primarily consists of several multi-choice clarification questions $A_t=\{Q_{t1},Q_{t2},\cdots,Q_{tn}\}$\footnote{We set $n=3$ in our implementation.}. Each question includes the content of the question itself and a set of answer candidates. It is important to note that, in each turn, the agent outputs both the questions and the retrieved items as timely feedback, facilitating a customer-friendly application interface.

The task presents several key challenges:
1) Generating non-trivial clarification questions requires a significant amount of product knowledge.
2) The agent should possess the ability to understand user intentions accurately in order to retrieve the most relevant product items and generate dynamic and valuable clarification questions.
3) The establishment of a quantitative and automated evaluation protocol is necessary to facilitate the development of effective agents for this task.
This paper addresses the first two challenges in Section~\ref{sec:product_agent}, and the third challenge in Section~\ref{sec:evaluation}.

\section{\Agent}\label{sec:product_agent}

\subsection{Overview of~\Agent}
This section delineates the proposed solution~\Agent~for the introduced task.
As shown in Figure~\ref{fig:overview},~\Agent~executes conversational loops, each of which can be decomposed by three main steps. Before initializing the loops, the user commences the conversation by inputting a product category as a naive query.
We will provide an overview in this section and detailed implementation is discussed in Section~\ref{subsec:implementation}.

\paragraph{Category Analysis.}
To generate appropriate clarification questions that can determine the user demands efficiently, it is necessary to collect and summarize relevant and well-elaborated product knowledge based on known user demands of each turn.
To achieve this, the agent first synthesizes a query reflective of all the known user demands, and the query is used to retrieve relevant product items from databases. The retrieved items, which probably cover the desirable items of interest, are then summarized as statistics. These statistics encompass various product aspects and serve as a dynamic knowledge base that is constructed in real time based on the ongoing user demands, laying a foundation for the subsequent steps.

\paragraph{Item Search.}
We dynamically retrieve different relevant product items based on all the known demands. Specifically,~\Agent~first generates a natural language query based on the current demands by harnessing the Query Generation tool. The generated query is then used to retrieve the most suitable items that meet the user demands.

\paragraph{Clarification Question Generation.}
In this step, the agent poses several new clarification questions along with their answer candidates which are carefully curated by incorporating the knowledgeable statistics and known user demands.
The user is then prompted to efficiently clarify their requirements by either selecting options or providing additional personal information that reflects their interests in the interaction interface. This user feedback is valuable for the task as it helps improve the accuracy of the product search process.

The above steps are repeated in each turn of the dialogue, with the aim of making the user demands in subsequent turns more precise compared to previous turns. Consequently, the accuracy of the product search steadily improves over time.

\subsection{Key Implementation}\label{subsec:implementation}

\begin{table*}[tbp!]
\setlength{\tabcolsep}{7pt}
\centering
\scalebox{0.80}{
% \begin{tabular}{@{~~}l@{~~}l@{~~}r@{~~}r@{~~}r@{~~}}
\begin{tabular}{clcc}
\toprule

\textbf{Tool}               &  \textbf{Description}  
&  \textbf{Input}           &  \textbf{Output}

\\ \midrule

\textbf{Text2SQL}           &  Generate SQL query
&  Demands                  &  SQL Query
\\

\textbf{Category Analyze}   &  Summarize a certain category
&  Product items            &  Category statistics
\\

\textbf{Query Generation}   &  Generate NL query
&  Demands                  &  NL query
\\

\textbf{Retriever}          &  Retrieve items relevant to demands
&  NL query                 &  Product items
\\

\textbf{Question Generation}        &  Generate clarification questions
&  Demands + Category statistics    &  Clarification questions
\\

\bottomrule
\end{tabular}}
\caption{Available tools of~\Agent.}
\label{tab:tools}
\end{table*}

Our proposed~\Agent~consists of three key components which work collectively to solve the product demand clarification task.

\paragraph{Databases.}
We store all product items in both SQL and dense vector datasets, allowing for flexible retrieval strategies. As shown in Figure~\ref{fig:overview}, we employ SQL retrieval in the stage 1 since it enables us to efficiently retrieve items through exact matching. On the other hand, we leverage non-SQL retrieval in the stage 2 because non-SQL retrievers can return product items ordered by their relevance to the user demands, which cannot be typically accomplished by SQL retrieval.

\paragraph{Tools.}
We define 5 tools for~\Agent, which are listed in Table~\ref{tab:tools}. To enable SQL retrieval,~\Agent~uses the Text2SQL tool. This tool prompts LLMs with the history demands to generate practical SQL queries. The synthesized SQL query is then used to retrieve product items relevant to the current known demands. Then, the Category Analyzing tool is performed to generate summarized statistics. For non-SQL retrieval, we create the Query Generation tool, which generates concise natural language queries. Additionally, in stage 3, we utilize the Question Generation tool to fabricate reasonable clarification questions based on the summarized statistics and the user demands. To ensure the desired output format, we specify the json description for~\Agent.

It should be kindly noted that for the purpose of autonomous evaluation, the demand clarification task has been slightly simplified and hence does not require a tool router, which is typically used to select the tool for the next action.

\paragraph{Memory.}
All the user demands, including clarification questions and corresponding answers, are stored as structured objects in the memory module. When the agent receives responses from users, it extracts the answers of clarification questions posed in the previous turn. The Q\&A pairs are then saved as structured demand objects. Additionally, the memory module stores the dialogue records and searching history. When invoking certain tools, we integrate only the required information selected from the memory into the designated slots within the prompt.
The memory are continuously updated as the dialogue progresses, ensuring that the intelligent agent can effectively handle dynamically user requirements. The design is crucial as it enables the agent to recall user preferences, thereby providing personalized assistance.

\section{The~\Benchmark~Benchmark}\label{sec:evaluation}

Most existing information retrieval benchmarks focus on static search processes~\cite{qiu-etal-2022-dureader,long2022multi} that do not involve human-AI interactions, which may limit their effectiveness in evaluating practical and interactive performance of search systems. Therefore, with the aim of comprehensively investigating the performance of conversational e-commercial search agents in an automatic manner, we introduce~\Benchmark~, a benchmarking framework that evaluates retrieval performance in two distinct settings, namely the~\textit{traditional} setting (Section~\ref{subsec:traditional}) and the~\textit{conversational} (Section~\ref{subsec:conversational}) setting. Both~\Agent~and~\Benchmark~are implemented using the open-sourced framework~\textit{LlamaIndex}.

\begin{table}[tbp!]
\renewcommand{\arraystretch}{1.1}
\renewcommand{\tabcolsep}{5pt}
\centering
\scalebox{0.70}{
\begin{tabular}{lcccc}
\toprule

% \textbf{Setting}  &  \textbf{Documents}  &  \textbf{Category}  &  \textbf{Turn}  &  \textbf{Query}  &  \textbf{Len. of Query}  \\

\textbf{Setting}  &  \textbf{Documents}  &  \textbf{Query}  &  \textbf{Len. of Query}  \\

\midrule

\textbf{Trad.}  &  1,000,000  &  2,000   &  27.02  \\
\textbf{Conv.}  &  1,000,000  &  10,000  &  8.59 / 16.45 / 27.40 / 37.33 / 45.03  \\

\bottomrule

\end{tabular}}
\caption{
Statistics of the~\Benchmark~benchmark. We report the average length of queries generated by GPT-4.
% For the conversational setting, queries are dynamically created by~\Agent~in each turn, and thus the queries in this setting are five times as many as those in the traditional setting.
}
\label{tab:statistics}
\end{table}

\subsection{Document Set}
We begin by creating a extensive collection of documents (product items), which serves as the knowledge base for our benchmark. This collection consists of 1,000,000 documents across 20 categories. To construct this dataset, we utilize AliMe KG~\cite{li2020alimekg}, a domain knowledge graph in the field of e-commerce that supports the top-50 main categories on the Alibaba e-commerce platform. From each category, we randomly sample 50,000 documents, resulting in the dataset used for our benchmark. To ensure efficient data management, we perform Named Entity Recognition (NER) pre-processing~\cite{wang-etal-2021-improving} on all the data. This results in a unified feature definition across various product categories, enabling us to store all the documents in a relational SQL dataset. For detailed statistics and description of the dataset, please refer to Table~\ref{tab:statistics} and Appendix~\ref{app:ner}.

\subsection{Traditional Information Retrieval}
\label{subsec:traditional}

In the traditional setting of~\Benchmark, we evaluate the retrieval performance of the conventional product search task using the methodology employed in previous work~\cite{long2022multi,qiu-etal-2022-dureader}. It is important to note that this experimental setting, which serves as a counterpart to the conversation setting introduced in Section~\ref{sec:task},~\textit{does not} involve the use of the agent.
To construct queries used for evaluation, we randomly sample 100 documents per category from the entire document set, resulting in a total of 2000 documents. We then leverage the~\textit{Doc2Query} technique~\cite{gospodinov2023doc2query,alaofi2023can} to transform these selected documents into synthesized queries by prompting GPT-4. Subsequently, we measure the retrieval performance of different retrieval methods based on the synthesized queries. The prompt for Doc2Query and examples of synthesized queries can be found in Appendix~\ref{app:prompts}.

\subsection{Retrieval Strategies}
Our experiments employ three widely-adopted retrievers: BM25~\cite{robertson2009probabilistic}, General Text Embedding (GTE)\footnote{\url{https://help.aliyun.com/zh/dashscope/developer-reference/text-embedding-quick-start}}, and CoROM~\cite{long2022multi}. Besides, we also incorporate the techniques of reranking and fusion into the retrievers. The detailed descriptions are provided in Appendix~\ref{app:retrievers}.

\subsection{Conversational Information Retrieval}\label{subsec:conversational}
The performance of~\Agent~can be comprehensively explored in the conversational setting, where its effectiveness is manifested directly via the retrieval performance across dialogue turns.

\paragraph{User simulation.}
It is well-acknowledged that the nature of dialogues, including the high cost of dialogue annotation~\cite{schatzmann2005quantitative,keizer2010parameter}, poses a challenge for the efficient evaluation of dialogue systems, particularly in an end-to-end manner. To handle this issue, we harness LLMs as intelligent user simulators, thereby accomplishing the automatic evaluation of~\Agent. Specifically, a LLM-driven user simulator is presented with several clarification questions along with their corresponding candidates in each turn, and then is prompted to answer the questions with access to the ground truth item. It is worth noticing that to avoid potential information leakage, which may create shortcuts for the agent, we restrict the user simulator to only answer the given questions with provided candidates, thus simulating a more realistic and reliable situation.

For each queried document, we simulate 10 turns (5 user turns and 5 agent turns) with the user simulator starting the conversation by inputting the desired product category. Examples of simulated dialogues and the prompt for user simulation can be found in Appendix~\ref{app:prompts}.

\subsection{Automatic evaluation metrics}
Drawn from the evaluation methodology adopted in earlier information retrieval benchmarks~\cite{long2022multi}, we evaluate the retrieval performance in both settings using Mean Reciprocal Rank at 10 documents (MRR@10) and hit rate at 10 documents (HIT@10). For the reranking results, we only report the result of the MRR@10 metric since the hit rates are identical to their counterparts.

\section{Experiments}\label{sec:experiments}

\subsection{Results of Traditional Setting}
The overall experimental results of the traditional setting are reported in Table~\ref{tab:traditional}, from which we can make the following conclusions.

\begin{table}[tbp!]
\renewcommand{\arraystretch}{1.0}
\renewcommand{\tabcolsep}{6pt}
\centering
\scalebox{0.75}{
\begin{tabular}{lccc}
\toprule

\multirow{2}{*}{\bf Retriever}  &  \multicolumn{2}{c}{\bf Retrieve}  &  \bf Rerank  \\

\cmidrule(l{2pt}r{2pt}){2-3} \cmidrule(l{2pt}r{2pt}){4-4} 

&  \textbf{HIT@10}  &  \textbf{MRR@10}  &  \textbf{MRR@10}  \\

\midrule

\textbf{BM25}           &  34.80  &  26.18  &  33.35  \\

\textbf{GTE}            &  \bf{69.00}  &  \bf{52.29}  &  \bf{63.82}  \\

\textbf{CoROM}          &  61.65  &  44.95  &  57.41  \\

\textbf{BM25 + GTE}     &  37.70  &  26.69  &  35.81  \\

\textbf{BM25 + CoROM}   &  37.35  &  26.77  &  35.65  \\

\textbf{GTE + CoROM}    &  59.45  &  26.91  &  55.90  \\

\bottomrule

\end{tabular}}
\caption{Retrieval performance of the traditional setting.}
\label{tab:traditional}
\end{table}

\paragraph{Dense retrievers outperform BM25.}
In order to generate diverse literal expressions in the process of synthesizing queries, we deploy LLMs. This, however, presents a challenge to sparse retrievers such as BM25. Dense retrievers such as GTE and CoROM, which leverage semantic matching, enhance retrieval performance on the MRR@10 scores by notable 26.11\% and 18.11\%, respectively. It should be noted that the efficiency of the retrievers might be affected by the Doc2Query prompt in our attempt to simulate realistic user queries. Furthermore, we observe that CoROM, despite being trained on the e-commerce dataset, falls behind GTE – trained on generic datasets – in performance. We suspect that this is because GTE, due to its pre-training and contrastive fine-tuning stages, is better equipped to capture complex semantics.

\paragraph{Reranking can significantly improve the MRR scores.}
The results indicate that reranking can greatly boost the MRR@10 scores for all retrievers and retriever combinations by 7.17\%, 11.53\%, and 12.46\% for BM25, GTE, and CoROM, respectively. It also shows similar trends for retriever fusion.

\paragraph{Retriever fusion decreases both HIT@10 and MRR@10 scores.}
We have also tested additional fusion strategies assisted by LlamaIndex in our initial experiments. The reciprocal rerank fusion algorithm outperforms others in most instances. This suggests that naive fusion strategies are unable to enhance retrieval performance in this case.

\begin{table}[tbp!]
\renewcommand{\arraystretch}{1.0}
\renewcommand{\tabcolsep}{5pt}
\centering
\scalebox{0.75}{
    \begin{tabular}{clccc}
    \toprule
    
    \multirow{2}{*}{\bf LLM}  &  \multirow{2}{*}{\bf Retriever}  &  \multicolumn{2}{c}{\bf Retrieve}  &  \bf Rerank  \\
    
    \cmidrule(l{2pt}r{2pt}){3-4} \cmidrule(l{2pt}r{2pt}){5-5} 
    
    &  &  \textbf{HIT@10}  &  \textbf{MRR@10}  &  \textbf{MRR@10}  \\
    
    \midrule
    
    \multirow{6}{*}{\bf GPT-3.5}
    
    &  \textbf{BM25}        &  \bf{35.04}  &  \bf{27.26}  &  \bf{25.69}  \\
    
    &  \textbf{GTE}         &  8.49   &  4.95   &  7.18   \\
    
    &  \textbf{CoROM}       &  12.48  &  7.96   &  10.79  \\
    
    &  \textbf{B + G}       &  32.35  &  16.12  &  25.21  \\
    
    &  \textbf{B + C}       &  32.54  &  17.66  &  25.27  \\
    
    &  \textbf{G + C}       &  7.61   &  4.21   &  6.67   \\

    \midrule

    \multirow{6}{*}{\bf GPT-4}

    &  \textbf{BM25}        &  \bf 39.48  &  \bf 32.00  &  \bf 30.20  \\
    
    &  \textbf{GTE}         &  8.27   &  4.92   &  7.17   \\
    
    &  \textbf{CoROM}       &  13.86  &  9.11   &  12.54  \\
    
    &  \textbf{B + G}       &  36.93  &  18.36  &  29.91  \\
    
    &  \textbf{B + C}       &  37.02  &  20.37  &  30.13  \\
    
    &  \textbf{G + C}       &  7.57   &  4.15   &  6.80   \\
    
    \midrule

    \multirow{6}{*}{\bf Qwen}

    &  \textbf{BM25}        &  \bf 31.58  &  \bf 25.24  &  \bf 24.85  \\
    
    &  \textbf{GTE}         &  16.45  &  10.56  &  13.82  \\
    
    &  \textbf{CoROM}       &  20.71  &  13.80  &  17.77  \\
    
    &  \textbf{B + G}       &  30.65  &  16.78  &  24.61  \\
    
    &  \textbf{B + C}       &  30.79  &  17.98  &  24.79  \\
    
    &  \textbf{G + C}       &  14.26  &  7.74   &  12.36  \\
    
    \bottomrule
    
    \end{tabular}
}
\caption{Retrieval performance of the conversational setting. We report results of fusion retrievers represented as X + Y, such as B + G (BM25 + GTE).}
\label{tab:conversational}
\end{table}

\subsection{Results of Conversational Setting}\label{subsec:results_conversational}
In the conversational setting of~\Benchmark, we employ OpenAI GPT-3.5-turbo-0125 to power the user simulator for all experiments, ensuring a fair comparison for~\Agent~built on various LLM backbones.
We benchmark a series of agents with Qwen-max-0107\footnote{\url{https://help.aliyun.com/zh/dashscope/developer-reference/quick-start}}, GPT-3.5-turbo-0125, and GPT-4-turbo-0409 as the backbones in this paper.

The results of conversational retrieval are detailed in Table~\ref{tab:conversational}, with several distinctions compared to the traditional setting.

\paragraph{BM25 surpasses dense retrievers.}
Surprisingly, dense retrievers, which excel in the traditional setting, yield significantly lower performance than BM25 across all LLMs. This can be attributed to the fact that synthesized queries are based on user responses, most of which are essentially provided choices of clarification questions without much modification. Consequently, the challenge faced by BM25 - computing relevance scores through exact term matching - is eliminated. This underlines the importance of generating queries with a specific pattern that is preferred by the utilized retriever.

\paragraph{Reranking boosts the MRR@10 scores, except those of BM25.}
We hypothesize that the level of noise in queries is so minimal that BM25 does not require the further application of an additional reranker. Other retrievers, however, struggle in capturing subtle semantics, leaving room for the reranker to augment their performance. In the same vein, the retriever fusion technique does \textit{not} enhance retrieval performance.

\begin{figure}[tbp!]
\centering
\includegraphics[scale=0.42]{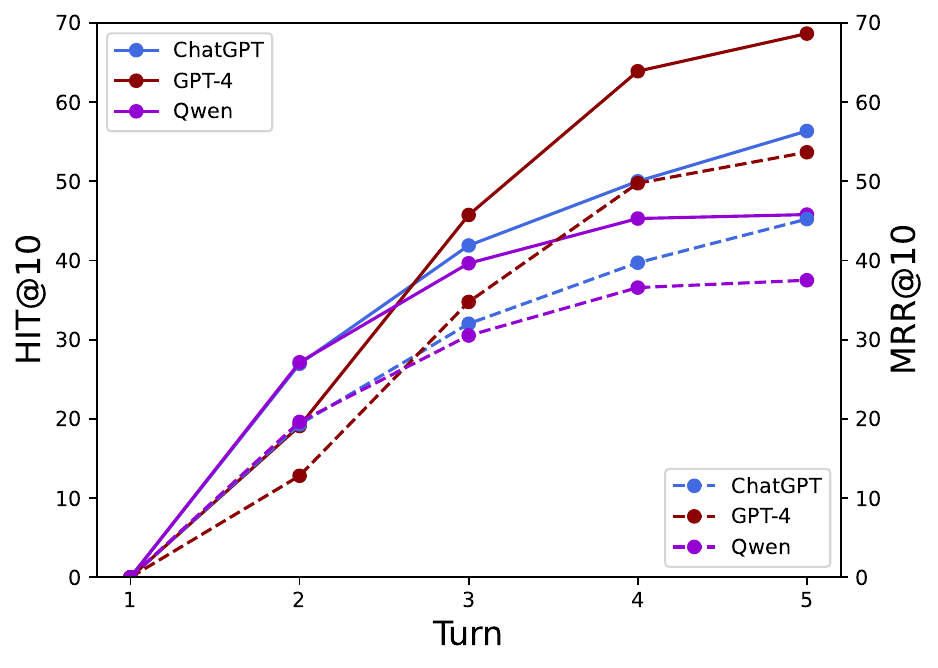}
\caption{Retrieval performance of~\Agent~driven by different LLMs with increasing turns using BM25.}
\label{fig:turn_bm25}
\end{figure}

\section{Analysis}

\subsection{Effect of Interaction Turns}
Figure~\ref{fig:turn_bm25} presents the trends in retrieval performance as interaction turns increase. The first turn can not retrieve any ground truth product items due to the access to only product categories. We observe that as the number of turns increase, all retrieval scores improve regardless of the types of retrievers employed, signifying the effectiveness of posing clarification questions. Notably, GPT-4 scores lower at the second turn but outperforms all the other LLMs after the third turn, which means GPT-4 tend to ask clarification questions from a global perspective.

\begin{figure}[tbp!]
\centering
\includegraphics[scale=0.42]{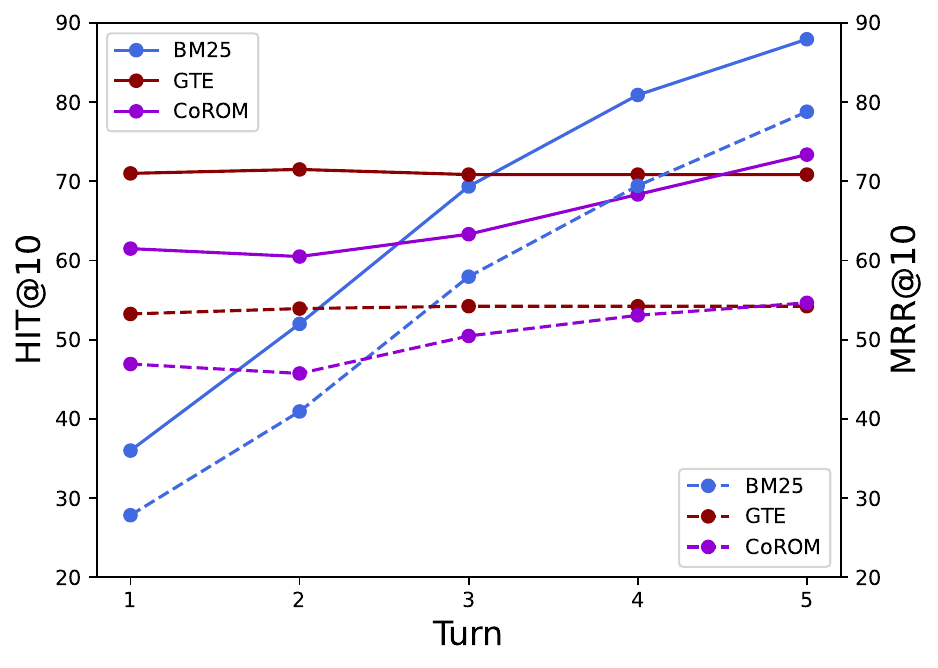}
\caption{Retrieval performance of the variant of the conversational setting. All scores are from~\Agent~driven by GPT-4.
}
\label{fig:turns_increase}
\end{figure}

\subsection{Performance Increase of Clarification}
To further demonstrate the effectiveness of clarification, we replace the initial user queries, i.e., product categories, with the synthesized queries introduced in Section~\ref{subsec:traditional}. The results illustrated in Figure~\ref{fig:turns_increase} indicate that retrievers benefit further from clarification, except for GTE. In detail, the performance of BM25 improves dramatically, similar to the results of conversational setting in Section~\ref{subsec:results_conversational}. However, the performance of GTE falls into a plateau. We speculate the reason is that general-purpose retrievers fail to capture subtle product features that are rare in their training data.

\subsection{Ablation Study}
Considering that summarized product statistics constitute a crucial part of the context for generating clarification questions, we perform a comprehensive ablation study focusing on strategies of acquiring statistics. This is conducted with the aim of probing the influence of dynamic statistics on~\Agent. Specifically, we experiment with three different strategies to derive the statistics:
1) for the ``w/o Statistics'' variant, we do not supply statistics to the process of clarification question generation.
2) For the ``Random'' variant, we randomly sample product items to summarize the statistics, ignoring dynamic user demands.
3) For the ``BM25'' variant, we substitute the SQL retriever with BM25, and
4) for the ``CoROM'' variant, similarly, we replace the SQL retriever with CoROM.

The ablation study results, which are shown in Table~\ref{tab:ablation}, reveal that BM25 and CoROM retrievers are better suited for statistic summarization. Additionally, both removing statistics or ignoring real-time user demands can impair the performance of product search. We analyze further this phenomenon in Section~\ref{subsec:failures}.

% A detailed analysis of why the SQL retriever performs worse than other retrievers is provided in Section~\ref{subsec:failures}.

\begin{table}[tb!]
\renewcommand{\arraystretch}{1.0}
\renewcommand{\tabcolsep}{6pt}
\centering
\scalebox{0.75}{
\begin{tabular}{lcc}
\toprule

\textbf{Retriever}  &  \textbf{HIT@10}  &  \textbf{MRR@10}  \\

\midrule

\textbf{w/o Statistics} &  15.60  &  10.69  \\

\textbf{Random}         &  39.50  &  19.54  \\

\textbf{BM25}           &  \bf 47.00  &  \bf 38.51  \\

\textbf{CoROM}          &  45.00  &  38.09  \\

\textbf{SQL} (Default)  &  39.90  &  32.40  \\

\bottomrule
\end{tabular}}
\caption{Ablation results of different strategies of acquiring statistics.}
\label{tab:ablation}
\end{table}

\begin{table}[tb!]
\renewcommand{\arraystretch}{1.0}
\renewcommand{\tabcolsep}{6pt}
\centering
\scalebox{0.75}{
\begin{tabular}{lcc}
\toprule

\bf{}  &  \bf{Invalid SQL} \%  &  \bf{Trivial SQL} \%  \\
\midrule

\textbf{GPT-3.5}        &  1.21   &  54.59  \\

\textbf{GPT-4}          &  3.52   &  55.36  \\

\textbf{Qwen-max}       &  3.06   &  44.92  \\

\bottomrule

\end{tabular}}
\caption{Failure rates of Text2SQL.}
\label{tab:failures}
\end{table}

\subsection{Failures of~\Agent}\label{subsec:failures}
We identify several failure types for~\Agent~throughout our experiments, all of which are due to the generation of either invalid or trivial SQL queries when using the Text2SQL tool.
% \footnote{Note that connection failures are not within the scope of our investigation.}
Specifically, invalid SQL queries mean Text2SQL generates queries that cannot be executed, while trivial SQL queries refer to those that retrieve nothing. The occurrence rates of both kinds of failures are presented in Table~\ref{tab:failures}, which reveals that all the involved LLMs have a tendency to produce trivial SQL queries. Delving deeper into the generated SQL queries, we discovered that trivial ones are typically produced in later turns, whereby LLMs form complex queries by merely amalgamating all known user demands. We addressed these issues by creating statistics derived from previously retrieved items, leading to~\Agent~sometimes failing to propose valuable questions directly related to the user's real-time demands. In the future, the exploration of more sophisticated SQL generation is worth investigating to enhance both \Agent's persona and effectiveness.

Additionally, we observe~\Agent~tends to generate clarification questions when conversations progress further. So we analyze the similarity of synthesized clarification questions, and the results are reported in Appendix~\ref{app:similarity}.

% \subsection{Case Study}

% \input{chapters/analysis}
\section{Conclusion}

This paper formulates the task of product demand clarification and presents the solution called~\Agent. In order to implement the evaluation in an automatic and quantitative manner , we introduce the~\Benchmark~benchmark with the aid of a user simulator.
Experiments indicate that~\Agent~significantly enhances the retrieval performance with increasing dialogue turns. We also identify certain limitations in the agent.

% , providing a pathway for potential enhancements to create more intelligent conversational agents in the future.

% The agent poses strategic clarification questions to pinpoint user demands by integrates three core modules, namely databases, memory and tools. 

% In order to implement the evaluation in an automatic and quantitative manner , we introduce the~\Benchmark~benchmark with the aid of a user simulator.
% Experiments indicate that~\Agent~significantly enhances the retrieval performance with increasing dialogue turns, regardless of the used backbone LLMs. We also identify certain limitations in the agent, providing a pathway for potential enhancements to create more intelligent conversational agents in the future.

\section*{Limitations}

\paragraph{Absence of human users.}
In this paper, we employ the user simulation technique to facilitate a more consistent and standardized evaluation process. Utilizing this approach allows for all the experiments conducted within our research to be completely void of any human involvement, contributing to cost-efficiency and saving time. We also acknowledge that potentially unidentified biases may arise from the hallucination effect of LLM-based user simulators within our evaluation pipeline.

\paragraph{Limitations of datasets.}
Our dataset was derived from the AliMe KG, encompassing multiple dimensions of products. Nonetheless, it does not provide comprehensive information that could be crucial to customers' purchase decisions, such as product pricing and customer reviews. Future research could delve into more diverse and higher-quality datasets to overcome these deficiencies.

\paragraph{Controllability of clarification questions.}
We encourage~\Agent~to ask clarification questions following in-context learning augmented with summarized statistics, without explicit and strong constraints. This strategy is not always optimal for all situations. In the future work, we will explore more effective planning strategies that can handle the task from a global perspective, providing a pathway for potential enhancements to create more intelligent conversational agents.

\paragraph{Lack of analysis of prompt sensitivity.}
The impact of prompt design on the performance of LLMs, often demonstrated as prompt sensitivity, is widely recognized. In our research, however, we did not conduct a systematic examination concerning the sensitivity of prompts used to develop the agent and user simulator. Detailed engineering of prompts could serve as a valuable measure for enhancing the performance of the~\Agent~in future studies.

\section*{Ethics Statement}
We are aware that our proposed~\Agent~may be potentially misused for improper purposes, such as privacy data collection and excessive propaganda. However, this vulnerability is not unique to our approach but a common threat to many LLM-based applications. It also highlights the significance of conducting appropriate regulations and enhancing the safety of LLMs in the future.

\bibliography{main}

\appendix

\section{Experimental Details}

\subsection{NER Pre-processing}\label{app:ner}
To accomplish the retrieval process in a consistent manner for different product categories, we initially apply named entity recognition (NER) to the documents obtained from AliMe KG. For this paper, we chose to use an e-commerce specific NER model~\cite{wang-etal-2021-improving} trained with a Cooperative Learning objective. This coaching strategy encourages two input views to generate similar contextual representations or output label distributions. The NER processing transform all the data from various categories into structured documents with a consistent named entity schema consisting of 54 hierarchical entity labels. We then compress this label space into a new one composed of only 10 entity labels, which allows the agent to retrieve documents more efficiently and accurately from the SQL database. The details of the product item description are provided in Table~\ref{tab:item_description}.

\subsection{Details about Retrievers}\label{app:retrievers}

\paragraph{BM25.}
As the representative of sparse retrievers, BM25~\cite{robertson2009probabilistic} consistently exhibits impressive performance across diverse information retrieval benchmarks~\cite{long2022multi}.

\paragraph{GTE.}
The General Text Embedding (GTE) model is a general-purpose text embedding model trained with the multi-stage contrastive learning objective~\cite{li2023towards}. It has shown exceptional results in the Massive Text Embedding Benchmark~\cite{muennighoff-etal-2023-mteb}.

\paragraph{CoROM.}
The CoROM model~\cite{long2022multi} uses BERT-base~\cite{devlin-etal-2019-bert} as its backbone. It is a dual Encoder-based embedding model~\cite{karpukhin-etal-2020-dense} trained on annotated query-passage e-commerce datasets.

\paragraph{Reranker.}
We also integrate the lightweight bge-reranker-base model~\cite{bge_embedding}\footnote{\url{https://huggingface.co/BAAI/bge-reranker-base}} to rerank the top-k documents retrieved.

\paragraph{Fusion.}
Furthermore, we also seek to investigate whether the fusion of multiple heterogeneous retrievers can boost performance on our benchmark. Consequently, we test hybrid fusion retrievers that utilize the reciprocal rerank fusion algorithm~\cite{cormack2009reciprocal} without requiring any additional models or significant computation.

\begin{table}[tb!]
\renewcommand{\arraystretch}{1.0}
\renewcommand{\tabcolsep}{4pt}
\centering
\scalebox{0.69}{
    \begin{tabular}{llc}
    \toprule
    
    \textbf{Name}   &  \textbf{Description}  &  \textbf{Type}  \\
    \midrule

    \textbf{Category}  &  Category of the item.  &  str  \\
    
    \textbf{Brand}  &  Brand of the item.  &  List[str]  \\
    
    \textbf{Series}  &  Series of the item.  &  List[str]  \\
    
    \textbf{Target Customer}  &  Target customers of the item.  &  List[str]  \\

    \textbf{Applicable Scenario}  &  Applicable scenarios of the item.  &  List[str]  \\

    \textbf{Decorative Attribute}  &  Decorative attributes of the item.  &  List[str]  \\

    \textbf{Material} &  Material of the item.  &  List[str]  \\

    \textbf{Style} &  Styles of the item.  &  List[str]  \\

    \textbf{Specification} &  Available specification of the item.  &  List[str]  \\

    \textbf{Color} &  Available colors of the item.  &  List[str]  \\

    \textbf{Function} &  Function of the item.  &  List[str]  \\

    \bottomrule
    \end{tabular}
}
\caption{Product item description.}
\label{tab:item_description}
\end{table}

\subsection{Similarity of Synthesized Clarification Questions}
\label{app:similarity}

We observe~\Agent~tends to generate clarification questions that overlap with those previously presented, which can negatively impact the task effectiveness and overall user satisfaction. Hence, we take a qualitative approach to measuring the similarity among clarification questions proposed in different dialogue turns by using BERTScore~\cite{Zhang*2020BERTScore:}.
BERTScore computes a similarity score for each token in the candidate sentence by comparing it to each token in the reference sentence. In the experiment, we concatenate the question content and its answer choices into a sentence. By treating all other questions and choices as references, We compute the similarity score for each synthesized clarification questions.

\begin{figure}[tbp!]
\centering
\includegraphics[scale=0.40]{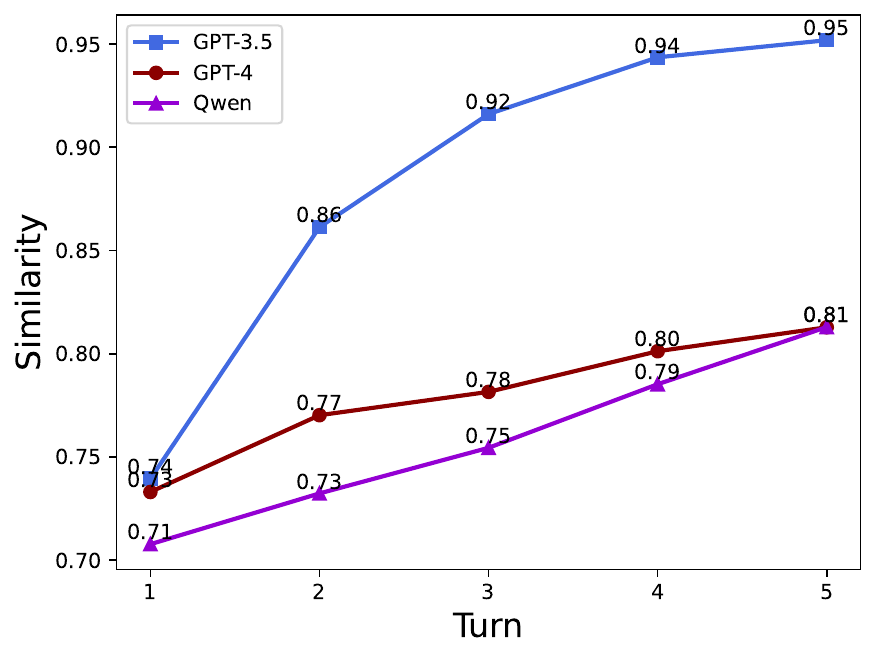}
\caption{Similarity of synthesized clarification questions computed by BERTScore.}
\label{fig:similarity}
\end{figure}

As Figure~\ref{fig:similarity} illustrates, we discern an upward trend in the similarity of synthesized clarification questions as dialogues progress. This suggests that although retrieval performance continues to enhance over ten dialogue turns, we may be still able to improve further the performance by boosting the diversity of clarification questions. Additionally, we observe that the agent powered by GPT-3.5 is more inclined to fabricate similar questions than those powered by GPT-4 and Qwen-max, despite the fact that the former achieves higher retrieval scores than the agent powered by Qwen-max when both utilize BM25 for item search.

\begin{figure*}[tbp!]
\centering
\includegraphics[scale=0.38]{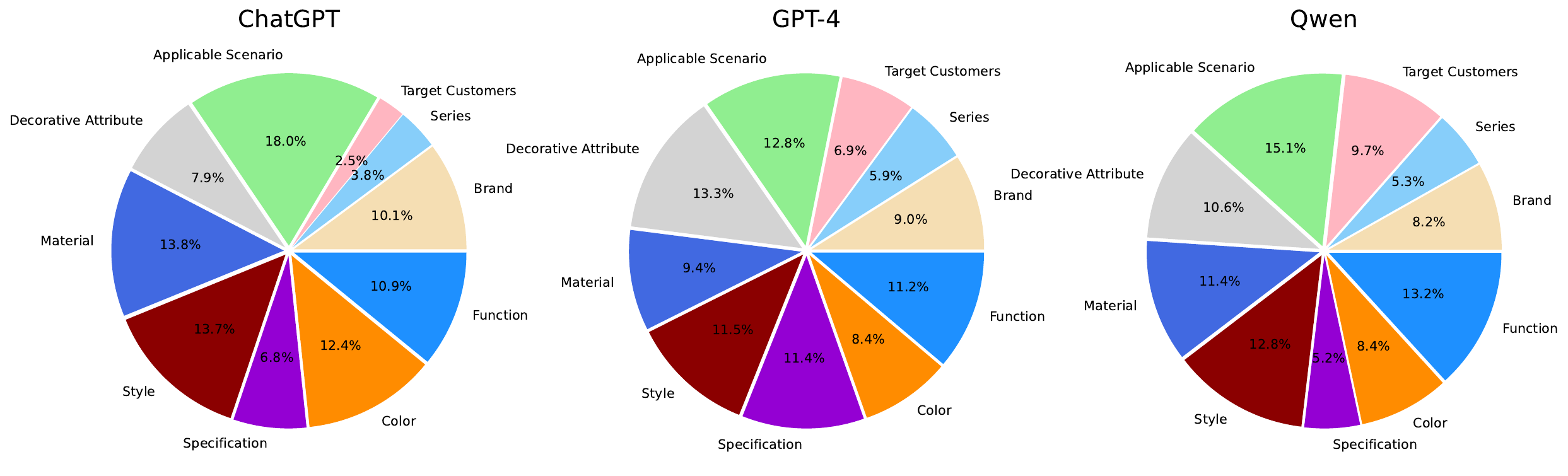}
\caption{Proportion of product aspects of clarification questions generated by three LLMs.}
\label{fig:aspects}
\end{figure*}

\subsection{Product Aspects of Clarification Questions}
We observe most clarification questions focus on one specific product aspects indicated in Table~\ref{tab:item_description}, such as category, brand, and series. Therefore, we analyze the proportion of product facets to which clarification questions are related. The results, illustrated in Figure~\ref{fig:aspects}, show that all three LLMs tend to generate clarification questions with even distributions. To detail, three aspects, namely applicable scenarios, style, function, respectively account for more than 10\% across LLMs. On the other hand, LLMs are less likely to clarify on series and specification, which may be too specialized for users. This reveals that LLMs naturally have a preference for asking specific types clarification questions in this task, since we do not incorporate any explicit constrains on clarification question generation.

\section{Prompts and Examples}\label{app:prompts}

\subsection{Prompts for~\Agent}
We report all the prompts used in our proposed~\Agent, including Text2SQL in Table~\ref{tab:prompt_text2sql}, Query Generation in Table~\ref{tab:prompt_query_generation} and Clarification Question Generation in Table~\ref{tab:prompt_question_generation}.

\subsection{Prompt for User Simulator}
We list the prompt for user simulation in Table~\ref{tab:prompt_user_simulator}.

\subsection{Examples of Conversation}
Here we provide a detailed conversation example regarding ``Canvas shoes" in Table~\ref{tab:full_example}. We can notice that~\Agent~first generates some critical clarification questions, such as the color, applicable scenarios and functions, which are helpful to precisely identify the basic user demands. Subsequently, the agent probes for additional specifics like style, decorative attributes and material. On the other hand, the user simulator, which is driven by GPT-3.5, occasionally offers responses that go beyond the provided options.

\begin{table*}
    \renewcommand{\tabcolsep}{0pt}
    \renewcommand{\arraystretch}{1.0}
    \centering
    \resizebox{0.90\textwidth}{!}{
        \begin{tabular}{l}
        \toprule
        \textbf{Text2SQL Prompt Template}  \\
        \midrule
        \makecell{
            \multicolumn{1}{p{.92\textwidth}}{
                You are a SQL generation assistant. Given several constraints, you need to generate a syntactically correct \{dialect\} SQL query statement to retrieve target records. In order to generate reasonable queries, you must follow the following rules:
            }  \\ \\
            \multicolumn{1}{p{.92\textwidth}}{
                1. Directly generate only SQL query statements, without outputting any explanation or inference information.
            }  \\
            \multicolumn{1}{p{.92\textwidth}}{
                2. Directly use ``*'' to retrieve all columns.
            }  \\
            \multicolumn{1}{p{.92\textwidth}}{
                3. Use the keyword ``LIMIT'' to limit the maximum number of retrieved records to \{max\_number\}.
            }  \\
            \multicolumn{1}{p{.92\textwidth}}{
                4. Carefully construct the where conditions for the query and use the keyword 'LIKE' as much as possible.
            }  \\ \\
            \multicolumn{1}{p{.92\textwidth}}{
                Given the following table structure description, only retrieval of that table is allowed:
            }  \\
            \multicolumn{1}{p{.92\textwidth}}{
                \textbf{Table schema}: \{schema\}
            }  \\ \\
            \multicolumn{1}{p{.92\textwidth}}{
                \textbf{Example input}:
            }  \\
            \multicolumn{1}{p{.92\textwidth}}{
                \textbf{Product category}: Casual pants
            }  \\
            \multicolumn{1}{p{.92\textwidth}}{
                \textbf{Question}: What material is the main material for the casual pants you would like to purchase?
            }  \\
            \multicolumn{1}{p{.92\textwidth}}{
                \textbf{Answer}: Polyester fiber
            }  \\ \\
            \multicolumn{1}{p{.92\textwidth}}{
                \textbf{Example output}:
            }  \\
            \multicolumn{1}{p{.92\textwidth}}{
                \textbf{SQL Query}: SELECT * From item WHERE category='Casual pants' AND material LIKE '\% polyester fiber\%' LIMIT \{max\_number\};
            }  \\ \\
            \multicolumn{1}{p{.92\textwidth}}{
                \textbf{Input}:
            }  \\
            \multicolumn{1}{p{.92\textwidth}}{
                \textbf{Product category}: \{category\}
            }  \\
            \multicolumn{1}{p{.92\textwidth}}{
                \textbf{Question}: \{question\_1\}
            }  \\
            \multicolumn{1}{p{.92\textwidth}}{
                \textbf{Answer}: \{answer\_1\}
            }  \\
            \multicolumn{1}{p{.92\textwidth}}{
                ...
            }  \\ \\
            \multicolumn{1}{p{.92\textwidth}}{
                \textbf{SQL Query}:
            }  \\
        }  \\
        \bottomrule
        \end{tabular}
    }
\caption{The one-shot prompt template of the Text2SQL tool in~\Agent.}
\label{tab:prompt_text2sql}
\end{table*}

\begin{table*}
    \renewcommand{\tabcolsep}{0pt}
    \renewcommand{\arraystretch}{1.0}
    \centering
    \resizebox{0.90\textwidth}{!}{
        \begin{tabular}{ll}
        \toprule
        \textbf{Query Generation Prompt Template}  \\
        \midrule
        \makecell{
            \multicolumn{1}{p{.92\textwidth}}{
                You are a query generation assistant. Given the user's purchasing demands, you need to generate a short natural language query statement (Query) to retrieve the target product. In order to generate a reasonable query, you must follow the following rules:
            }  \\ \\
            \multicolumn{1}{p{.92\textwidth}}{
                1. The generated query should be concise, composed of keywords, and separated by spaces.
            }  \\
            \multicolumn{1}{p{.92\textwidth}}{
                2. The generated query should cover all of the user's purchasing requirements.
            }  \\
            \multicolumn{1}{p{.92\textwidth}}{
                3. Do not output any explanations or inference information, and do not use unnecessary punctuation such as quotation marks.
            }  \\ \\
            \multicolumn{1}{p{.92\textwidth}}{
                \textbf{Product category}: \{category\}
            }  \\
            \multicolumn{1}{p{.92\textwidth}}{
                \textbf{Question}: \{question\_1\}
            }  \\
            \multicolumn{1}{p{.92\textwidth}}{
                \textbf{Answer}: \{answer\_1\}
            }  \\
            \multicolumn{1}{p{.92\textwidth}}{
                ...
            }  \\ \\
            \multicolumn{1}{p{.92\textwidth}}{
                \textbf{Query}:
            }  \\
        }  \\
        \bottomrule
        \end{tabular}
    }
\caption{The zero-shot prompt template of the Query Generation tool in~\Agent.}
\label{tab:prompt_query_generation}
\end{table*}

\begin{table*}
    \renewcommand{\tabcolsep}{0pt}
    \renewcommand{\arraystretch}{1.0}
    \centering
    \resizebox{0.90\textwidth}{!}{
        \begin{tabular}{ll}
        \toprule
        \textbf{Clarification Question Generation Prompt Template}  \\
        \midrule
        \makecell{
            \multicolumn{1}{p{.92\textwidth}}{
                You are a product shopping assistant that can accurately identify user demands, and you are capable of generating three multiple-choice questions for demand clarification. To help you ask valuable questions, here is a summary of statistics about \{category\}.
            }  \\
            \multicolumn{1}{p{.92\textwidth}}{
                \textbf{Statistics}: \{statistics\} 
            }  \\ \\
            \multicolumn{1}{p{.92\textwidth}}{
                1. The generated content must focus on the product category (\{category\}) and contribute to accurately identifying user demands.
            }  \\
            \multicolumn{1}{p{.92\textwidth}}{
                2. It is prohibited to generate new questions that are duplicates of previous ones.
            }  \\
            \multicolumn{1}{p{.92\textwidth}}{
                3. When constructing options, try to directly select them from the statistical data.
            }  \\ \\
            \multicolumn{1}{p{.92\textwidth}}{
                You need to generate new multiple-choice questions for demand clarification based on the historical Q\&A. Here is the historical Q\&A:
            }  \\ \\
            \multicolumn{1}{p{.92\textwidth}}{
                \textbf{Question}: \{question\_1\}
            }  \\
            \multicolumn{1}{p{.92\textwidth}}{
                \textbf{Answer}: \{answer\_1\}
            }  \\
            \multicolumn{1}{p{.92\textwidth}}{
                ...
            }  \\ \\
            \multicolumn{1}{p{.92\textwidth}}{
                The generated result should strictly follow the JSON format, and no additional information should be included.
            }  \\
            \multicolumn{1}{p{.92\textwidth}}{
                \textbf{JSON format description}: {json\_description}
            }  \\
        }  \\ \\
        \multicolumn{1}{p{.92\textwidth}}{
           \`{}\`{}\`{}json
        }  \\
        \bottomrule
        \end{tabular}
    }
\caption{The zero-shot prompt template of the Clarification Question Generation tool in~\Agent. We require LLM to synthesize clarification questions in structured json form.}
\label{tab:prompt_question_generation}
\end{table*}

\begin{table*}
    \renewcommand{\tabcolsep}{0pt}
    \renewcommand{\arraystretch}{1.0}
    \centering
    \resizebox{0.90\textwidth}{!}{
        \begin{tabular}{ll}
        \toprule
        \textbf{User Simulator Prompt Template}  \\
        \midrule
        \makecell{
            \multicolumn{1}{p{.92\textwidth}}{
                You are in a conversation with a shopping assistant, hoping that they can help you search for suitable products. To provide reasonable answers, you must follow these rules:
            }  \\
            \multicolumn{1}{p{.92\textwidth}}{
                1. You must answer the question accurately based on the target product.
            }  \\
            \multicolumn{1}{p{.92\textwidth}}{
                2. Try to use the options provided for your answer, if your answer is not covered by the options, simply answer "Other".
            }  \\
            \multicolumn{1}{p{.92\textwidth}}{
                3. Directly output the answer, do not provide any explanation or reasoning information and avoid unnecessary punctuation like quotation marks.
            }  \\
            \multicolumn{1}{p{.92\textwidth}}{
                4. Each question's answer should be on a separate line.
            }  \\ \\
            \multicolumn{1}{p{.92\textwidth}}{
                Here is the target product information: \{item\}
            }  \\ \\
            \multicolumn{1}{p{.92\textwidth}}{
                Here is the official question from the assistant: \{questions\}
            }  \\ \\
            \multicolumn{1}{p{.92\textwidth}}{
                Answer the above question directly, do not repeat the question, **each answer should be on a separate line**.
            }  \\
        }  \\
        \bottomrule
        \end{tabular}
    }
\caption{The zero-shot prompt template of the user simulator.}
\label{tab:prompt_user_simulator}
\end{table*}

\begin{table*}[tbh!]
    \renewcommand{\tabcolsep}{0pt}
    \renewcommand{\arraystretch}{1.0}
    \centering
    \resizebox{0.98\textwidth}{!}{
        \begin{tabularx}{\linewidth}{X}
        \toprule
        \textbf{Conversation example by GPT-4 driven~\Agent~and the user simulator}  \\
        \midrule
            \textbf{A}: I am a product sales assistant. Please just type in the targeted product category first, and I would search for you. \\ \\
            \textbf{U}: Canvas shoes.  \\ \\
            
            \textbf{A}: I have search several relevant items for you: <Item$_{1}$>, <Item$_{2}$>, $\cdots$. Also, you can answer these questions so that I could recommend more suitable items for you! \\
            
            \textbf{Q1}: When you need canvas shoes, which color do you mainly focus on? \\
            \textbf{C1}: ["Light gray", "Sky blue", "Yellow", "Military Green", "Rose Red", "Other"] \\
            
            \textbf{Q2}: Which applicable scenarios do you prefer when wearing canvas shoes? \\
            \textbf{C2}: ["Travel", "Dance", "Fitness", "Construction Site", "Official Flagship", "Other"] \\
            
            \textbf{Q3}: When choosing canvas shoes, what functions do you value? \\
            \textbf{C3}: ["Anti slip", "Increase by 10CM", "Waterproof bottom", "Anti puncture", "Antibacterial", "Other"] \\ \\

            \textbf{U}:
            \textbf{Answers for Q1}: "Military Green". \\
            \textbf{Answers for Q2}: "Travel". \\
            \textbf{Answers for Q3}: "Other". \\ \\

            \textbf{A}: I have search several relevant items for you: <Item$_{1}$>, <Item$_{2}$>, $\cdots$. Also, you can answer these questions so that I could recommend more suitable items for you! \\

            \textbf{Q1}: When choosing canvas shoes, which style do you prefer? \\
            \textbf{C1}: ["Leisure", "Internal elevation", "Ties", "Flat Heel", "Low help", "Other"] \\
            
            \textbf{Q2}: What upper decoration attributes do you usually pay attention to? \\
            \textbf{C2}: ["Adhesive", "Breathable", "Inside", "Mesh surface", "Sole", "Other"] \\
            
            \textbf{Q3}: When purchasing canvas shoes, which material do you prefer to make them from? \\
            \textbf{C3}: ["Mesh", "Rubber", "Canvas", "Other"] \\ \\

            \textbf{U}:
            \textbf{Answers for Q1}: "Leisure". \\
            \textbf{Answers for Q2}: Coffee leopard print, leopard print, round head. \\
            \textbf{Answers for Q3}: "Canvas". \\ \\

            \textbf{A}: I have search several relevant items for you: <Item$_{1}$>, <Item$_{2}$>, $\cdots$. Also, you can answer these questions so that I could recommend more suitable items for you! \\

            ... \\
            
        \midrule

        \textbf{Ground truth product item}  \\

        \midrule

            \textbf{Title}: Feiyue thick-soled canvas shoes for women, spring 2023, leopard print loafers, casual and versatile mantou shoes 8925

            \textbf{Category}: Canvas shoes,
            \textbf{Brand}: ["Feiyue"],
            \textbf{Series}: [],
            \textbf{Target Customer}: ["Female", "18-40 years old", "Youth"],
            \textbf{Applicable Scenario}: ["Spring"],
            \textbf{Decorative Attribute}: ["Shoe bottom", "Thick-soled", "Cross straps", "Coffee leopard print", "Shallow mouth", "Leopard print", "Pattern", "Round head", "Khaki leopard print"],
            \textbf{Material}: ["Rubber", "Canvas"],
            \textbf{Style}: ["Lace-up", "Low-cut", "Leopard print", "Casual", "Bread", "Steamed bun", "All-match", "Vulcanization process"]
            \textbf{Specification}: ["EUR34", "EUR35", "EUR36", "EUR37", "EUR38", ...]
            \textbf{Color}: ["Purple", "Lake blue and orange", "Army green and white", "Pink", "Mixed colors"]
            \textbf{Function}: []  \\

        \bottomrule
        
        \end{tabularx}
    }
\caption{A conversation example between~\Agent~(A) and the user (U). We also list the ground truth product item that the user is searching for.}
\label{tab:full_example}
\end{table*}

% item_703360912217

% \section{Additional Results}

% \subsection{}

\end{document}